\documentclass[aps,pre,twocolumn,groupedaddress,showpacs,floatfix]{revtex4}
\usepackage{amsmath}
\usepackage{amssymb}
\usepackage{graphicx}

\begin{document}

\title{Beyond the average: detecting global singular nodes
from local features in complex networks}

\author{Luciano da Fontoura Costa} 
\affiliation{Instituto de F\'{\i}sica de S\~ao Carlos. 
Universidade de S\~ ao Paulo, S\~{a}o Carlos, SP, PO Box 369,
13560-970, phone +55 16 3373 9858,FAX +55 16 3371 3616, Brazil,
luciano@if.sc.usp.br}
\author{Marcus Kaiser}
\affiliation{School of Computing Science, University of 
Newcastle, Claremont Tower, Newcastle upon Tyne, NE1 7RU, United
Kingdom. \\  Also: Institute of Neuroscience, Henry Wellcome Building for
Neuroecology, University of Newcastle, Framlington Place, Newcastle
upon Tyne, NE2 4HH, United Kingdom}
\author{Claus Hilgetag}
\affiliation{International University Bremen, School of Engineering 
and Science, Campus Ring 6, 28759 Bremen, Germany. \\ Also: Boston
University, Sargent College, Department of Health Sciences, 635
Commonwealth Ave Boston, MA 02215, USA }

\date{20th July 2006}

\begin{abstract}   
Deviations from the average can provide valuable insights about the
organization of natural systems.  This article extends this important
principle to the more systematic identification and analysis of
singular local connectivity patterns in complex networks.  Four
measurements quantifying different and complementary features of the
connectivity around each node are calculated and multivariate
statistical methods are then applied in order to identify outliers.
The potential of the presented concepts and methodology is illustrated
with respect to a word association network.
\end{abstract}

\maketitle

\emph{`Everything great and intelligent is in the minority'}
(J. W. von Goethe)
\vspace{0.5cm}

While uniformity and regularity are important properties of patterns
in nature and science, it is the \emph{minority deviations} in such
patterns which often are particularly informative.  A prototypical
example of such a fact is the great importance given by animal
perception to variations in signals, in detriment of constant stimuli.
For instance, the outlines of shapes/objects play a much more
important role in visual perception than uniform regions (see, for
instance~\cite{Marr:1980}).  The power of cartoons, involving only a
few contour lines, is an immediate consequence of this perceptual
rule.  At the same time, our focus of visual attention is frequently
driven by deviations cues at the visual periphery (e.g. a dot of
contrasting color, a small object movement or flashes) -- even during
saccadic eye movements -- i.e. abrupt, ballistic gaze displacements,
changes (e.g. a flash) in the scene can be
perceived~\cite{Kaiser_Lappe:2004}.

Many are the examples of the importance of minority in other
scientific areas, including mathematics (the importance of extremal
values) and physics (e.g. entropy).  In complex networks
(e.g.~\cite{Albert_Barab:2002,Newman:2003,Boccaletti:2006}), the
uniformity of connections is typically expressed with respect to the
number of connections of each node, the so-called \emph{node degree}.
Amongst the most uniformly connected types of networks are the random
networks -- also called \emph{Erd\H{o}s-R\'enyi (ER)}
networks~\cite{ErdosRenyi:1961}, characterized by constant probability
of connection between any pair of nodes.  Because of its uniformity,
the connectivity of this type of network can be well approximated in
terms of the average and standard deviation of their node degrees,
which is a consequence of its concentrated, Gaussian-like, degree
distribution (e.g.~\cite{Albert_Barab:2002}).  Despite being
understood in depth since the first half of the 20th century, ER
networks played a relatively minor role as a model of natural
phenomena.  Actually, it is rather difficult to find a natural model
which can be properly represented and modeled by the Poisson-based ER
networks. It was mainly through the investigations of sociologists
(e.g.~\cite{Milgram:1967}) and, more recently, the identification of
power law distributions of node degree in the
Internet~\cite{Faloutsos99} and WWW (e.g.~\cite{Albert_Barab:2002}),
that complex networks became widely known.  The success of complex
networks stems mainly from the fact that a large and representative
range of structured and heterogeneous natural and human-made systems
have been found to fall into this category.  The importance of
deviations was therefore once again testified.

While global deviation from uniformity was ultimately the reason
behind the success of complex networks, a good deal of attention has
been focused in identifying uniformities in complex networks, such as
in node degree distributions (e.g.~\cite{Albert_Barab:2002}).  While
such approaches are also important, only a relatively few works have
targeted local singularity identification.  For instance, Milo et
al.~\cite{Milo:2002} addressed the detection of motifs significantly
deviating from those in random networks (see
also~\cite{Sporns_Kotter:2004}), while Guimer\`a and
Amaral~\cite{Guimera:2005} investigated the special role of nodes at
the borders between communities (e.g.~\cite{Newman:2004}).

The methodology proposed in the current article includes two steps:
First, measurements~\cite{Costa_surv:2006} of the local connectivity
are obtained for each node; then, outlier detection methodologies from
multivariate statistics and pattern recognition
(e.g.~\cite{Johnson_Wichern:2002}) are applied in order to identify
the nodes exhibiting the greatest deviations.  The considered
measurements include the normalized average and coefficient of
variation of the degrees of the immediate neighbors of a node -- a
measurement related to the hierarchical node degree
(e.g.~\cite{Costa_PRL:2004,Costa_EPJ:2006, Costa_JSP:2006}), their
clustering coefficient (e.g.~\cite{Albert_Barab:2002}), and the
locality index, an extension of the matching index
(e.g.~\cite{Kaiser_Hilgetag:2004}) to consider all the immediate
neighbors of each node, instead of individual edges.

The article is organized as follows.  First, we present the basic
concepts in complex networks and the adopted measurements.  Then, the
proposed methodology for singularity detection is presented and its
potential is illustrated with respect to a word association network.
This specific experimental network has been specifically chosen
because of its potential non-homogeneity of connections and more
accessible interpretation of the results.

A non-directed complex network (or graph) is a discrete structure
defined as $\Gamma = \left( V, Q \right)$, where $V$ is a set of $N$
nodes and $Q$ is a set of $E$ non-directed edges.  Complex networks
can be effectively represented in terms of their respective
\emph{adjacency matrix} $K$, such that the presence of an undirected
link between nodes $i$ and $j$ is expressed as $K(i,j)=K(j,i)=1$.  The
\emph{degree} of any given node $i$ can be calculated as $k(i) =
\sum_{p=1}^{N} K(p,j)$.  Note that the node degree provides a simple and 
immediate quantification of the connectivity at the individual node
basis.  Nodes which have a particularly high degree (usually appearing
in minority), the so-called \emph{hubs}, are known to play a
particularly important role in the connectivity of complex networks
(e.g.~\cite{Albert_Barab:2002}).  For instance, they provide shortcuts
between the nodes to which they connect.  Other features of the local
connectivity of a network can be quantified by using several
measurements such as those adopted in the current work, which are
presented as follows.

{\bf Neighboring degree (normalized average and coefficient of
variation):} An alternative measurement which, though not frequently
used, provides valuable information about local connectivity is the
average and coefficient of variation of the \emph{neighboring degree}
of each node $i$.  By neighboring degree it is meant the set of
degrees of the immediate neighbors of $i$, excluding connections with
the reference node $i$. These two measurements are henceforth
abbreviated as $a(i)$ and $cv(i)$.  Note that the latter can be
obtained by dividing the standard deviation of the neighboring degrees
of node $i$ by the respective average.  The average neighboring degree
is closely related to the second hierarchical degree
(e.g.~\cite{Costa_PRL:2004,Costa_EPJ:2006,Costa_JSP:2006}), which
corresponds to the sum of the neighboring degrees.  Therefore, the
average neighboring degree of a node $i$ can be calculated by dividing
the second hierarchical degree by the number of immediate neighbors of
$i$.  Because the values of $a(i)$ tend to increase with the degree of
node $i$, we consider its normalized version $r(i)=a(i)/k(i)$.  The
measurement $cv(i)$ provides a natural quantification of the relative
variation of the connections established by the neighboring nodes. For
instance, in case all neighboring nodes exhibit the same number of
connections (i.e. degree), we have that $cv(i)=0$.  Values larger than
1 are typically understood as indicating significant variation.

{\bf Clustering coefficient:} This measurement, henceforth abbreviated
as $cc(i)$ is defined as follows: given a reference node $i$,
determine the number of edges between its immediate neighbors and
divide this number by the maximum possible number of such
connections. This traditional and widely used measurement
(e.g.~\cite{Albert_Barab:2002}) quantifies the degree in which the
neighbors of the reference node $i$ are interconnected, with $0 \leq
cc(i) \leq 1$.

{\bf Locality index:} This measurement has been motivated by the
\emph{matching index}~\cite{Kaiser_Hilgetag:2004}, which is
adapted here in order to reflect the `internality' of the connections
of all the immediate neighbors or a given reference node, instead of a
single edge.  More specifically, given a node $i$, its immediate
neighbors are identified and the number of links between themselves
(including the reference node, in order to avoid singularities at
nodes with unit degree) is expressed as $N_{int}(i)$ and the number of
connections they established with nodes in the remainder of the
network, including the reference node $i$, is expressed as
$N_{ext}(i)$.  The locality index of node $i$ is then calculated as
$loc(i) = N_{int}(i)/(N_{int}(i)+N_{ext}(i))$.  Note that $0 < loc(i)
\leq 1$.  In case all connections of the neighboring nodes are
established between themselves, we have that $loc(i) = 1$.  This value
converges towards zero as higher percentages of external connections
are established.

Note that the four measurements considered (i.e. $r(i)$, $cv(i)$,
$cc(i)$ and $loc(i)$) therefore provide objective and complementary
information about the local connectivity around each network node,
paving the way for effective identification of local singularities.  A
number of statistically-sound concepts and methods have been developed
which allow the identification of \emph{outliers} in data sets
(e.g.~\cite{Johnson_Wichern:2002}).  The detection of connectivity
singularities arising locally in complex networks can therefore be
approached in terms of the following two steps:

\newcounter{indx}
\begin{list}
    {(\roman{indx})}{\usecounter{indx} \setlength{\rightmargin}{\leftmargin}}
    \item  Map the local connectivity
    properties around each node, after quantification in terms of
    measurements such as those adopted in the current work,
    into a respective feature vector $\vec{X}$;  and 
    \item Detect the outliers, which are understood as local
    singularities of the network under analysis, in the respectively
    induced feature space.
\end{list}

In the present work, as we restrict our attention to four measurements
of local connectivity around each node, we have a 4-dimensional
feature space.  Each node is therefore mapped by the measurements into
4-dimensional vectors $\vec{X}$ which `live' in the 4-dimensional
feature space, defining distributions of points in this space.  In
order to facilitate visualization, such dispersions of points can be
projected onto the plane by using the principal component analysis
methodology (e.g.~\cite{Johnson_Wichern:2002,Costa_Sporns:2005,
Costa_surv:2006}).  First, the covariance matrix $\Sigma$ of the data
is estimated and the eigenvectors corresponding to the largest
absolute eigenvalues are calculated and used to project the cloud of
points into a space of reduced dimensionality.  It can be shown that
this methodology ensures the concentration of variance along the first
main axes.

The identification of outliers represents an important subject in
multivariate analysis and pattern
recognition(e.g.~\cite{Johnson_Wichern:2002,Duda_Hart:2001}).
Basically, \emph{outliers} are instances of the observations which are
particularly different.  Because no formal definition of outlier
exists, one of the most traditional and effective means for their
identification~\cite{Johnson_Wichern:2002} relies on the \emph{visual}
inspection of the data distribution in feature spaces: outliers would
be the points which are further away from the main concentration of
data in the feature space.  Because such distributions can be skewed
and elongated, comparisons with the center of mass of the data is
often unsuitable. A quantitative
methodology~\cite{Johnson_Wichern:2002} which allows for more general,
Gaussian-like, multivariate distributions is to use the Mahalanobis
distance.  So, outliers are identified as corresponding to the feature
vectors $\vec{X}$ implying particularly large values of the
\emph{Mahalanobis distance}, defined as
\begin{equation}
  D(X) = \sqrt{(\vec{X}-\vec{\mu})^T \Sigma^{-1} (\vec{X}-\vec{\mu})},
\end{equation}
where $T$ stands for matrix transposition, $\vec{\mu}$ is the average
of the feature vectors and $\Sigma$ is the respective covariance
matrix.

Note that the latter method works in the original 4-dimensional space
and therefore requires no data projections.  Except for too high
dimensional feature spaces or intricate, concave feature
distributions, these two methods tend to produce congruent results.

Before proceeding to the illustration of the suggested methodology for
identification of singularities, it is worth discussing briefly what
could be the origin of such deviations in complex networks.  For the
sake of clarity, we organize and discuss the main sources of
singularity according to the following four major categories:

{\bf Growth Dynamics:} The most natural and direct origin of
singularities is that they are a consequence of the own network growth
dynamics.  An important example of such a phenomenon is the appearance
of hubs in scale free networks.  However, many other types of dynamics
may lead to singularities, especially when growth is affected by the
dynamics undergone by the network and the dynamics itself involves
singularities.

{\bf Community structure:} Several complex networks contain a number
of communities which, as discussed elsewhere
(e.g.~\cite{Guimera:2005}), imply different roles for nodes.  For
example, nodes which are at the borders of the community tend to
connect to nodes both in its respective community as well as to a few
nodes in other communities.  

{\bf Parent node influence:} In the common case where the network
supports a dynamical process (e.g. Internet, WWW, protein-protein
interaction, among many others), it is possible that singular dynamics
taking place at a specific node ends up by influencing its immediate
neighborhood.  For instance, in case of social networks, one
individual may convince its immediate acquaintances to assume specific
behavior.  As a simple example, the parent node may convince its
friends that they should seek reclusion, in which case their
respective node degrees would tend to become small, implying low
neighboring degree.  Similar effects can be characterized in many
other types of networks.  

{\bf External influences:} Singularities may also arise as a
consequence of factors which are external to the network.  For
instance, in a geographical network, it is possible that some of its
nodes be located in a region promoting different connectivity.  As a
simple example, in flight routes networks, localities inside a
particularly rich region tend to have more interconnected flights,
increasing the neighboring degree.

In order to illustrate the potential of the singularity identification
procedure with respect to real networks, we considered the word
association data obtained through psychophysical experiments described
in~\cite{Costa_what:2004,Costa_PRL:2004}.  In this experiment, whose
objective is to map pairwise associations between words, a single
initial word (`sun') is presented by the computer to the subject, who
is required to reply with the first word which comes to his/her mind.
Except for the first word, all others are supplied by the subject.
This procedure minimizes the streaming of associations which could be
otherwise implied. Networks are obtained from such associations by
considering each word as a node and each association as an edge.
Because of the rich structure of word associations, which suggests
power law degree distributions~\cite{Costa_what:2004}, such a network
favors the appearance of singularities of local connectivity.  In
addition, its non-specialized nature allows an intuitive and simple
discussion of the detected singularities.  The relatively small size
of this network, which involves $N=302$ nodes and $E=854$ edges, also
facilitates the illustration of the combined use of feature space
visualization and Mahalanobis distance.  The originally weighted
network, with the weights given by the frequency of associations, was
thresholded (i.e. any link with non-zero weight was considered as an
edge) and symmetrized (i.e. $K = \delta(K,K^T)$, where $\delta$ is the
Kronecker delta applied in elementwise fashion).

\begin{figure}[h]
 \begin{center} 
   \includegraphics[scale=0.4,angle=0]{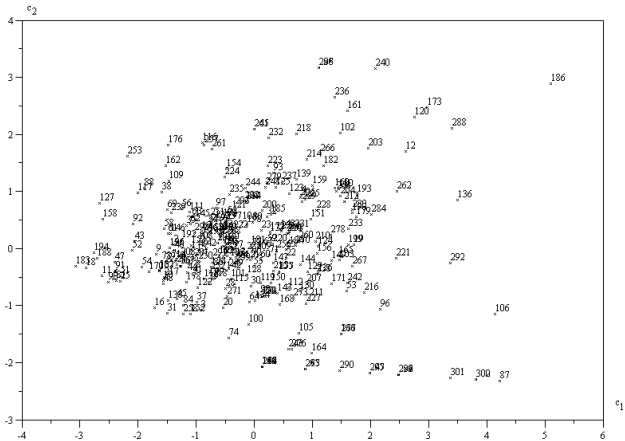} \\
   \caption{The feature space obtained by principal component projections
   of the four dimensional measurement vectors of the word
   association network.~\label{fig:outliers}}
\end{center}
\end{figure}

Figure~\ref{fig:outliers} shows the feature space obtained after
principal components projection of the 4-dimensional feature space
into the plane.  In order to remove scaling bias, the four adopted
measurement were standardized (e.g.~\cite{Johnson_Wichern:2002})
before principal component analysis projection.  Each of the axes
corresponds to linear combinations of the 4 original measurements,
more specifically, $c_1 =0.69r-0.12cv+0.21cc-0.68loc$ and
$c_2=-0.05r-0.76cv+0.65cc-0.02loc$, which indicates that all
measurements contributed significantly to the projection.

The twelve most singular nodes (i.e. words), corresponding to the
respectively largest values of the Mahalanobis distances (also
considering previous standardization of the measurements), are shown
in decreasing order in Table~\ref{tab:Mah}, where the first three rows
include the overall average, minimum and maximum values of the
respective features.

\begin{table}
  \vspace{1cm} 
\begin{tabular}{||l|r|r|r|r|r|r||} \hline 
$Word$       & $k$   & $a$   &  $r$  & $cv$  & $cc$ & $loc$ \\ \hline 
average      &  5.66 & 9.12  & 2.38  & 0.60  & 0.17 & 0.38 \\ \hline
minimum      &  1    & 2.00  & 0.18  & 0.00  & 0.00 & 0.09 \\ \hline
maximum      &  35   & 21.50 & 10.75 & 1.40  & 1.00 & 0.88 \\ \hline
183 (good)   &  35   & 6.43  & 0.18  & 0.57  & 0.05 & 0.88 \\ 
186 (land)   &  2    & 21.50 & 10.75 & 0.89  & 1.00 & 0.09 \\ 
106 (breath) &  2    & 20.50 & 10.25 & 0.45  & 0.00 & 0.09 \\ 
18 (long)    &  27   & 6.19  & 0.23  & 0.57  & 0.04 & 0.84 \\ 
136 (service)&  2    & 19.00 & 9.50  & 1.19  & 0.00 & 0.10 \\ 
87 (saddle)  &  1    & 10.00 & 10.00 & 0.00  & 0.00 & 0.10 \\ 
300 (sharp)  &  1    & 9.00  & 9.00  & 0.00  & 0.00 & 0.10 \\ 
302 (bear)   &  1    & 9.00  & 9.00  & 0.00  & 0.00 & 0.10 \\ 
194 (two)    &  24   & 7.08  & 0.30  & 0.64  & 0.07 & 0.81 \\ 
292 (finger) &  2    & 17.50 & 8.75  & 0.77  & 0.00 & 0.10 \\ 
241 (ask)    &  2    & 3.00  & 1.50  & 0.47  & 1.00 & 0.50 \\ 
265 (answer) &  2    & 3.00  & 1.50  & 0.47  & 1.00 & 0.50 \\ \hline 
\end{tabular} 
\caption{The twelve most singular nodes obtained for
  the word association network and their respective non-normalized
  features.}~\label{tab:Mah}
\end{table}

For the sake of complementing the following discussion, the
traditional node degrees and the non-normalized average neighboring
degrees are also given in the first two columns, respectively.  In
addition, many of the extremal (i.e. minimum and maximum) values of
each feature are present among the detected singularities.  The
singularities in Table~\ref{tab:Mah} can be divided into groups of
words as follows:

{\bf Group 1 (183, 18, 194):} These singular words are characterized
by relatively high values of locality index, high node degree
(i.e. they are hubs), medium values of $r$ and low values of $cc$.
Such properties indicate that these words are associated to many
others words which are not in the immediate neighborhood.  These three
words appear at the left-hand extremity of the data distribution in
Figure~\ref{fig:outliers}.  Interestingly, they correspond to `good',
`long' and `one', which are adjectives.

{\bf Group 2 (186):} This word not only has connectivity features
which are different from all other words in Table~\ref{tab:Mah}, but
also appears particularly isolated in the feature space (upper
right-hand corner) in Figure~\ref{fig:outliers}.  It is characterized
by low degree but high neighboring degree, reflected in the highest
relative neighboring degree value (10.75).  It also exhibits a high
coefficient of variation and maximum clustering coefficient, while the
locality index is particularly low (the minimum for the network).
Therefore, this word has been associated to two other words which
present markedly distinct degrees and which are themselves
interrelated.  Not surprisingly, those two words are the common
adjectives `good' and `bad', with respective degrees of 35 and 8.  In
this sense the measurement $cv$ is capable of expressing the asymmetry
of the connections established by the immediate neighbors. This word
is best understood as a second hierarchical level hub.

{\bf Group 3 (106, 136, 292):} These three words are similar to that
in Group 2 (and, as that word, are also substantives), except that they
present lower clustering coefficient (i.e. the two immediate neighbors
are not interconnected) and coefficient of variation (i.e. the degrees
of the immediate neighbors are more uniform).  These three words can
be found at the lower right-hand corner of the feature space in
Figure~\ref{fig:outliers}.

{\bf Group 4 (87, 300, 302):} These words have unit degree, therefore
exhibiting low $cv$, $cc$ and $loc$.  These words were not exercised
particularly during the experiment because they appeared near its
conclusion.  They can be found at the extremity of the alignment of
points in the feature space in Figure~\ref{fig:outliers},
corresponding to other words with similar properties.  Note that the
own aligned group of cases is itself a mesoscopic singularity of the
network.

{\bf Group 5 (241, 265):} These two related verbs, `ask' and `answer',
are characterized by having two immediate neighbors, each of them
being interconnected and establishing two connections with other
network nodes.  Interestingly, the obtained symmetry of local
connections reflected the inherently symmetry of these two words.

An additional interesting point follows that, from a theoretical point
of view, each feature could have 2 kinds of outliers: those towards
the minimum and those towards the maximum of that particular
feature. For 4 features, there are thus $2^4=16$ possible groups of
outliers. Looking at the word association, we find 5 groups of main
outliers. Why are some groups of outliers found whereas 11 potential
groups are absent? Among the several possible reasons we have that:
(a) some groups are absent (skewed feature distribution) (b) some
groups are present but not included in the top 12 singularities (c)
some features strongly correlate with each other leading to the merger
of potential outlier groups. For example, if a minimum feature A
correlates with a maximum in feature B (negative correlation),
outliers may form one group AB. However, if all features are
statistically independent and distributions are non-skewed, all
potential groups of outliers should also occur in the top list.  In
short, looking at absent outlier groups (a singularity of the
singularity pattern) can provide additional information about the
nature of the network connectivity.

The above results, which could by no means be inferred from the visual
inspection of the network, illustrate the effectiviness and
complementariness of the four adopted measurements in providing the
basis for sound singularity detection, with good agreement between the
Mahalanobis values and the distribution in the projected feature
space.  A series of peculiar local connectivity features were
identified which allowed interesting interpretations.

It is not by chance that hubs and communities have become particularly
important in complex network research: they correspond to structural
singularities.  In this work we extended the general principle that
minority deviations are essential in order to analyze the local
connectivity around each node in a network.  Four complementary
measurements, all stable to small perturbations
(e.g.~\cite{Costa_surv:2006}), have been used to derive 4-dimensional
informative feature vectors.  Two multivariate methodologies,
including visualization after standardization and principal component
projections, as well as the calculation of the Mahalanobis distances
in the full feature space, have been applied in order to identify the
twelve most singular nodes in the word association network, which were
divided into five main groups presenting distinctive properties.  We
are currently applying the suggested methodology to a number of other
important real networks, with similarly encouraging results.  Possible
future works include the consideration of broader context around each
node (e.g. by using the hierarchical schemes described
in~\cite{Costa_PRL:2004,Costa_EPJ:2006,Costa_JSP:2006}) as well as the
application of the method for the analysis of each detected
community. Another promising work would be to consider singularity
identification during network growth or dismantling
(e.g. attacks). More sophisticated alternatives for outlier detection
are also possible, especially by using hierarchical clustering
algorithms (e.g.~\cite{Duda_Hart:2001,Costa_JSP:2006}) in order to
obtain further information about how the singularities fit in the
overall network structure.

\vspace{1cm}

Luciano da F. Costa is grateful to CNPq (308231/03-1) and FAPESP for
financial support.

\bibliographystyle{apsrev}
\bibliography{singul_words}

\begin{thebibliography}{21}
\expandafter\ifx\csname natexlab\endcsname\relax\def\natexlab#1{#1}\fi
\expandafter\ifx\csname bibnamefont\endcsname\relax
  \def\bibnamefont#1{#1}\fi
\expandafter\ifx\csname bibfnamefont\endcsname\relax
  \def\bibfnamefont#1{#1}\fi
\expandafter\ifx\csname citenamefont\endcsname\relax
  \def\citenamefont#1{#1}\fi
\expandafter\ifx\csname url\endcsname\relax
  \def\url#1{\texttt{#1}}\fi
\expandafter\ifx\csname urlprefix\endcsname\relax\def\urlprefix{URL }\fi
\providecommand{\bibinfo}[2]{#2}
\providecommand{\eprint}[2][]{\url{#2}}

\bibitem[{\citenamefont{Marr}(1980)}]{Marr:1980}
\bibinfo{author}{\bibfnamefont{D.}~\bibnamefont{Marr}},
  \emph{\bibinfo{title}{Vision}} (\bibinfo{publisher}{Freeman},
  \bibinfo{year}{1980}).

\bibitem[{\citenamefont{Kaiser and Lappe}(2004)}]{Kaiser_Lappe:2004}
\bibinfo{author}{\bibfnamefont{M.}~\bibnamefont{Kaiser}} \bibnamefont{and}
  \bibinfo{author}{\bibfnamefont{M.}~\bibnamefont{Lappe}},
  \bibinfo{journal}{Neuron} \textbf{\bibinfo{volume}{41}}, \bibinfo{pages}{293}
  (\bibinfo{year}{2004}).

\bibitem[{\citenamefont{Albert and Barab\'asi}(2002)}]{Albert_Barab:2002}
\bibinfo{author}{\bibfnamefont{R.}~\bibnamefont{Albert}} \bibnamefont{and}
  \bibinfo{author}{\bibfnamefont{A.~L.} \bibnamefont{Barab\'asi}},
  \bibinfo{journal}{Rev. Mod. Phys.} \textbf{\bibinfo{volume}{74}},
  \bibinfo{pages}{47} (\bibinfo{year}{2002}).

\bibitem[{\citenamefont{Newman}(2003)}]{Newman:2003}
\bibinfo{author}{\bibfnamefont{M.~E.~J.} \bibnamefont{Newman}},
  \bibinfo{journal}{SIAM Review} \textbf{\bibinfo{volume}{45}},
  \bibinfo{pages}{167} (\bibinfo{year}{2003}),
  \bibinfo{note}{cond-mat/0303516}.

\bibitem[{\citenamefont{Boccaletti et~al.}(2006)\citenamefont{Boccaletti,
  Latora, Moreno, Chavez, and Hwang}}]{Boccaletti:2006}
\bibinfo{author}{\bibfnamefont{S.}~\bibnamefont{Boccaletti}},
  \bibinfo{author}{\bibfnamefont{V.}~\bibnamefont{Latora}},
  \bibinfo{author}{\bibfnamefont{Y.}~\bibnamefont{Moreno}},
  \bibinfo{author}{\bibfnamefont{M.}~\bibnamefont{Chavez}}, \bibnamefont{and}
  \bibinfo{author}{\bibfnamefont{D.-U.} \bibnamefont{Hwang}},
  \bibinfo{journal}{Physics Reports} \textbf{\bibinfo{volume}{424}},
  \bibinfo{pages}{175} (\bibinfo{year}{2006}),
  \bibinfo{note}{cond-mat/0303516}.

\bibitem[{\citenamefont{Erd{\H{o}}s and R{\'{e}}nyi}(1961)}]{ErdosRenyi:1961}
\bibinfo{author}{\bibfnamefont{P.}~\bibnamefont{Erd{\H{o}}s}} \bibnamefont{and}
  \bibinfo{author}{\bibfnamefont{A.}~\bibnamefont{R{\'{e}}nyi}},
  \bibinfo{journal}{Acta Mathematica Scientia Hungary}
  \textbf{\bibinfo{volume}{12}}, \bibinfo{pages}{261} (\bibinfo{year}{1961}).

\bibitem[{\citenamefont{Milgram}(1967)}]{Milgram:1967}
\bibinfo{author}{\bibfnamefont{S.}~\bibnamefont{Milgram}},
  \bibinfo{journal}{Psychology Today} pp. \bibinfo{pages}{61--67}
  (\bibinfo{year}{1967}).

\bibitem[{\citenamefont{Faloutsos et~al.}(1999)\citenamefont{Faloutsos,
  Faloutsos, and Faloutsos}}]{Faloutsos99}
\bibinfo{author}{\bibfnamefont{M.}~\bibnamefont{Faloutsos}},
  \bibinfo{author}{\bibfnamefont{P.}~\bibnamefont{Faloutsos}},
  \bibnamefont{and}
  \bibinfo{author}{\bibfnamefont{C.}~\bibnamefont{Faloutsos}},
  \bibinfo{journal}{Computer Communication Review}
  \textbf{\bibinfo{volume}{29}}, \bibinfo{pages}{251} (\bibinfo{year}{1999}).

\bibitem[{\citenamefont{Milo et~al.}(2002)\citenamefont{Milo, Shen-Orr,
  Itzkovitz, Kashtan, Chklovskii, and Alon}}]{Milo:2002}
\bibinfo{author}{\bibfnamefont{R.}~\bibnamefont{Milo}},
  \bibinfo{author}{\bibfnamefont{S.}~\bibnamefont{Shen-Orr}},
  \bibinfo{author}{\bibfnamefont{S.}~\bibnamefont{Itzkovitz}},
  \bibinfo{author}{\bibfnamefont{N.}~\bibnamefont{Kashtan}},
  \bibinfo{author}{\bibfnamefont{D.}~\bibnamefont{Chklovskii}},
  \bibnamefont{and} \bibinfo{author}{\bibfnamefont{U.}~\bibnamefont{Alon}},
  \bibinfo{journal}{Science} \textbf{\bibinfo{volume}{298}},
  \bibinfo{pages}{824} (\bibinfo{year}{2002}).

\bibitem[{\citenamefont{Sporns and K{\"{o}}tter}(2004)}]{Sporns_Kotter:2004}
\bibinfo{author}{\bibfnamefont{O.}~\bibnamefont{Sporns}} \bibnamefont{and}
  \bibinfo{author}{\bibfnamefont{R.}~\bibnamefont{K{\"{o}}tter}},
  \bibinfo{journal}{PLoS Biology 2} \textbf{\bibinfo{volume}{e369}}
  (\bibinfo{year}{2004}).

\bibitem[{\citenamefont{Guimer\`a and Amaral}(2005)}]{Guimera:2005}
\bibinfo{author}{\bibfnamefont{R.}~\bibnamefont{Guimer\`a}} \bibnamefont{and}
  \bibinfo{author}{\bibfnamefont{L.~A.~N.} \bibnamefont{Amaral}},
  \bibinfo{journal}{Nature} \textbf{\bibinfo{volume}{433}},
  \bibinfo{pages}{895} (\bibinfo{year}{2005}).

\bibitem[{\citenamefont{Newman}(2004)}]{Newman:2004}
\bibinfo{author}{\bibfnamefont{M.~E.~J.} \bibnamefont{Newman}},
  \bibinfo{journal}{Phys. Rev. E} \textbf{\bibinfo{volume}{69}},
  \bibinfo{pages}{1} (\bibinfo{year}{2004}).

\bibitem[{\citenamefont{da~F.~Costa et~al.}(2006)\citenamefont{da~F.~Costa,
  Rodrigues, Travieso, and Boas}}]{Costa_surv:2006}
\bibinfo{author}{\bibfnamefont{L.}~\bibnamefont{da~F.~Costa}},
  \bibinfo{author}{\bibfnamefont{F.~A.} \bibnamefont{Rodrigues}},
  \bibinfo{author}{\bibfnamefont{G.}~\bibnamefont{Travieso}}, \bibnamefont{and}
  \bibinfo{author}{\bibfnamefont{P.~R.~V.} \bibnamefont{Boas}}
  (\bibinfo{year}{2006}), \bibinfo{note}{cond-mat/0505185}.

\bibitem[{\citenamefont{Johnson and Wichern}(2002)}]{Johnson_Wichern:2002}
\bibinfo{author}{\bibfnamefont{R.~A.} \bibnamefont{Johnson}} \bibnamefont{and}
  \bibinfo{author}{\bibfnamefont{D.~W.} \bibnamefont{Wichern}},
  \emph{\bibinfo{title}{Applied multivariate statistical analysis}}
  (\bibinfo{publisher}{Prentice-Hall}, \bibinfo{year}{2002}).

\bibitem[{\citenamefont{da~F.~Costa}(2004{\natexlab{a}})}]{Costa_PRL:2004}
\bibinfo{author}{\bibfnamefont{L.}~\bibnamefont{da~F.~Costa}},
  \bibinfo{journal}{Phys. Rev. Letts.} \textbf{\bibinfo{volume}{93}},
  \bibinfo{pages}{098702} (\bibinfo{year}{2004}{\natexlab{a}}).

\bibitem[{\citenamefont{da~F.~Costa and da~Rocha}(2006)}]{Costa_EPJ:2006}
\bibinfo{author}{\bibfnamefont{L.}~\bibnamefont{da~F.~Costa}} \bibnamefont{and}
  \bibinfo{author}{\bibfnamefont{L.~E.~C.} \bibnamefont{da~Rocha}},
  \bibinfo{journal}{Eur. Phys. J. B} \textbf{\bibinfo{volume}{50}},
  \bibinfo{pages}{237} (\bibinfo{year}{2006}).

\bibitem[{\citenamefont{da~F.~Costa and Silva}(2006)}]{Costa_JSP:2006}
\bibinfo{author}{\bibfnamefont{L.}~\bibnamefont{da~F.~Costa}} \bibnamefont{and}
  \bibinfo{author}{\bibfnamefont{F.~N.} \bibnamefont{Silva}},
  \bibinfo{journal}{J. Stat. Phys.}  (\bibinfo{year}{2006}), \bibinfo{note}{in
  press, cond-mat/0412761}.

\bibitem[{\citenamefont{Kaiser and Hilgetag}(2004)}]{Kaiser_Hilgetag:2004}
\bibinfo{author}{\bibfnamefont{M.}~\bibnamefont{Kaiser}} \bibnamefont{and}
  \bibinfo{author}{\bibfnamefont{C.}~\bibnamefont{Hilgetag}},
  \bibinfo{journal}{Biol. Cybern.} \textbf{\bibinfo{volume}{90}},
  \bibinfo{pages}{311} (\bibinfo{year}{2004}).

\bibitem[{\citenamefont{da~F.~Costa and Sporns}(2005)}]{Costa_Sporns:2005}
\bibinfo{author}{\bibfnamefont{L.}~\bibnamefont{da~F.~Costa}} \bibnamefont{and}
  \bibinfo{author}{\bibfnamefont{O.}~\bibnamefont{Sporns}},
  \bibinfo{journal}{Eur. Phys. J. B} \textbf{\bibinfo{volume}{48}},
  \bibinfo{pages}{567} (\bibinfo{year}{2005}).

\bibitem[{\citenamefont{Duda et~al.}(2001)\citenamefont{Duda, Hart, and
  Stork}}]{Duda_Hart:2001}
\bibinfo{author}{\bibfnamefont{R.~O.} \bibnamefont{Duda}},
  \bibinfo{author}{\bibfnamefont{P.~E.} \bibnamefont{Hart}}, \bibnamefont{and}
  \bibinfo{author}{\bibfnamefont{D.~G.} \bibnamefont{Stork}},
  \emph{\bibinfo{title}{Pattern Classification}} (\bibinfo{publisher}{Wiley
  Interscience}, \bibinfo{year}{2001}).

\bibitem[{\citenamefont{da~F.~Costa}(2004{\natexlab{b}})}]{Costa_what:2004}
\bibinfo{author}{\bibfnamefont{L.}~\bibnamefont{da~F.~Costa}},
  \bibinfo{journal}{Intl. J. Mod. Phys. C} \textbf{\bibinfo{volume}{15}},
  \bibinfo{pages}{371} (\bibinfo{year}{2004}{\natexlab{b}}).

\end{thebibliography}
\end{document}